\documentclass[aps,pra,reprint,amsmath,amssymb,superscriptaddress,longbibliography]{revtex4-2} % single-spaced double column
\usepackage{color,graphicx,float,hyperref,stackengine,braket,bm,mathtools,calc,float,eqnarray}
\usepackage[caption=false]{subfig}
\begin{document}
\title{Static synthetic gauge field control of double optomechanically induced transparency in a closed-contour interaction scheme}
\author{Beyza \surname{S\"{u}tl\"{u}o\u{g}lu}}
\author{Ceyhun \surname{Bulutay}}
\email{bulutay@fen.bilkent.edu.tr}
\affiliation{Department of Physics, Bilkent University, Ankara 06800, Turkey}
\date{\today}
\begin{abstract}
We study theoretically an optical cavity and a parity-time ($\mathcal{PT}$)-symmetric pair of mechanical resonators, where all oscillators are pairwise coupled, forming an optomechanical system with a closed-contour interaction. Due to the presence of both gain and feedback, we explore its stability and the root loci over a wide coupling range. Under the red-sideband pumping and for the so-called $\mathcal{PT}$-unbroken phase it displays a double optomechanically induced transparency (OMIT) for an experimentally realizable parameter set.
We show that both the transmission amplitude and the group delay can be continuously steered from the lower transmission window to the upper one by the loop coupling phase which breaks the time-reversal symmetry and introduces a static synthetic gauge field.
In the $\mathcal{PT}$-unbroken phase both the gain-bandwidth and delay-bandwidth products remain constant over the full range of the controlling phase. Tunability in transmission and bandwidth still prevails in the $\mathcal{PT}$-broken phase, albeit over a reduced range. In essence, we suggest a simple scheme that grants coupling phase-dependent control of the single and double OMIT phenomena within an effective $\mathcal{PT}$-symmetric optomechanical system.
\end{abstract}

% PhySH
% Research Areas / Optomechanics, Quantum Optics, Light propagation, transmission & absorption
% Physical Systems / Quantum cavities

\maketitle

\section{Introduction}
% Literature Survey
Electromagnetically induced transparency is one of the striking quantum interference effects in light-matter interaction \cite{harris1990,boller1991}, which alters the optical susceptibility of a medium, leading to phenomena such as the giant Kerr effect, canceling the absorption, and the group delay or advance of a probe light around a specific wavelength \cite{fleischhauer2005}. By augmenting its basic level scheme it can support a double transparency window \cite{rebic2004polarization,joshi2005phase} which greatly enhances the nonlinear coupling between two copropagating weak signals \cite{wang2006large,macrae2008matched}. Its offspring in cavity optomechanics by harnessing the dispersively coupled radiation pressure encompasses optomechanically induced transparency (OMIT) \cite{agarwal2010,weis2010optomechanically,teufel2011,safavi2011,chang2011multistability} and double OMIT \cite{qu2013phonon,tarhan2013,shahidani2013control,ma2014tunable,wang2014optomechanical}.

Recently, the basic OMIT framework has been enriched in various directions. One of them was by introducing three-mode coupling, which gives rise to profound consequences. For instance, in the case of two photonic modes coupled to a mechanical resonator, switching from transparency to absorption by adjusting the strength of the cavity coupling \cite{lei2015three} and nonreciprocal amplification or attenuation have been reported \cite{li2017optical,xu2018}. This work was extended to reconﬁgurable
nonreciprocal transmission between two microwave modes \cite{bernier2017nonreciprocal} and nonreciprocal enhancement of second-order sidebands \cite{li2020}.
In a cavity-magnon system that utilizes pathway interference, switching between fast and slow light transmission has been proposed \cite{ullah2020pra} and experimentally achieved by tuning the relative phase of the magnon pumping and cavity probe \cite{zhao2021}. 
Other important progress has been brought about by parity-time ($\mathcal{PT}$)-symmetric 
photonic concepts that widen the range of opportunities \cite{ozdemir2019parity,elganainy2018}. For instance, under a varying gain-to-loss ratio, inverted OMIT is displayed, as well as the possibility to exchange slow and fast lights \cite{jing2015optomechanically,liu2017,lu2018}. In the quantum regime, for blue-detuned driving, the $\mathcal{PT}$ symmetry enables the elimination of the dissipation effect \cite{li2017theoretical}. In a recent study, within the three-mode paradigm the transition of the system from the so-called $\mathcal{PT}$-broken phase to the $\mathcal{PT}$-unbroken phase is accompanied by single to double OMIT, when the so-called exceptional point (EP) is crossed \cite{wang2019mechanical}. Indeed, this is one example of the extensive research efforts that have been dedicated to the interesting features of the EP in optomechanics, with a few others being nonreciprocal energy transfer between two eigenmodes of a mechanical system \cite{xu2016}, loss-induced transparency \cite{zhang2018loss}, enhanced sensitivity \cite{lau2018}, and sideband generation \cite{he2019}. These are complemented by other OMIT studies in systems having photonic-sector $\mathcal{PT}$ symmetry \cite{li2016parity,zhang2018double,jiang2018tunable,lu2019tunable}. 

One further direction to engineer electromagnetically induced transparency is by completing the basic $\Lambda$ scheme to a $\Delta$ coupling \cite{korsunsky1999phase,joshi2009phase,joo2010electromagnetically} 
which was used earlier for coherence population trapping \cite{kosachiov1991coherent}.
Very recently, this is utilized for nonreciprocal ground-state cooling \cite{lai2020nonreciprocal} and OMIT tunability by controlling the so-called dark-mode effect \cite{lai2020tunable}. It basically makes up a closed-contour interaction that embodies a synthetic gauge field which has totally revamped photonics \cite{onoda2004,raghu2008,haldane2008,fang2012realizing}. Going beyond the so-called static gauge field, in cases where the field freely evolves and behaves like a limit-cycle oscillator, it can acquire a dynamical degree of freedom of its own \cite{walter2016,zapletal2019A}. Both static and dynamical synthetic gauge fields have been employed in optomechanics \cite{fang2012photonic,hafezi2012,schmidt2015,walter2016,fang2017generalized,zapletal2019B,mathew2020,xu2020} and other hybrid systems involving atoms \cite{restrepo2017} and spins \cite{barfuss2018phase}. 

% This work
In this work, bringing together these concepts, we study theoretically an optical cavity coupled to a $\mathcal{PT}$-symmetric pair of mechanical resonators under a red-detuned pumping within the sideband-resolved regime. In our scheme all oscillators are pairwise coupled, in contrast to Ref.~\cite{wang2019mechanical}, thus comprising a closed-contour interaction \cite{barfuss2018phase}. This simple extension enables us to take advantage of the aforementioned advances. Because of the gain and feedback in the closed loop, first we work out its stability over the parameter space and the representative root loci. This clearly displays how their character drastically changes from the $\mathcal{PT}$-broken phase to the $\mathcal{PT}$-unbroken phase. The most conspicuous outcome is the switch from single to double OMIT behavior. This can be obtained in the stable region of an experimentally realizable parameter space. We show that both the transmission amplitude and slow light group delay of each transparency window can be continuously steered by any of the coupling phases in the closed-contour interaction, acting as a static synthetic gauge field \cite{fang2017generalized}. Both the gain-bandwidth and the delay-bandwidth products remain fairly constant over the full span of the coupling phase. In the $\mathcal{PT}$-broken phase an OMIT bandwidth tunability of about 50\% is provided..

% Plan
The paper is organized as follows. In Sec.~II we present our model, and its theoretical analysis for the probe transmission characteristics. In Sec.~III we discuss the parameter set we use for our calculations. Section~IV contains the results, starting with the stability analysis and followed by the control of the OMIT behavior, bandwidth, and slow light properties. Section V addresses the experimental relevance of our model, and our main conclusions are highlighted in Sec.~VI. Appendix~A introduces the gauge transformation that leads to the closed-loop phase starting from a general case; Appendix~B presents some analytical expressions derived by means of the so-called adiabatic elimination technique.
\begin{figure}[H]
  \centering
  \includegraphics[width=0.5\textwidth]{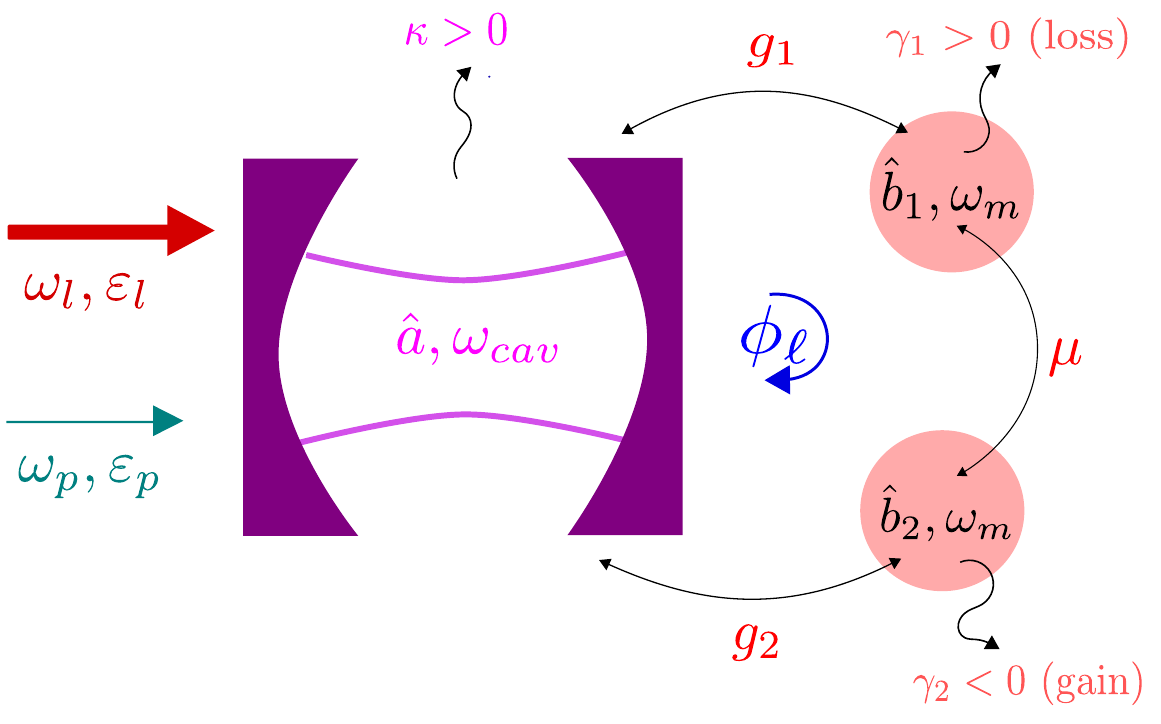}
\caption{Closed-contour interaction optomechanical system composed of a photonic cavity with the relevant resonance at $\omega_{cav}$, and two mechanical resonators with identical frequencies, $\omega_m$. Loss and gain rates are indicated with wavy arrows. A strong pump laser and a weak probe laser with angular frequencies $\omega_l$ and $\omega_p$, respectively, are externally coupled to the cavity.}
\label{fig1}
\end{figure}
\section{Theory}
We consider a ternary-coupled  system consisting of a photonic cavity attached to a pair of $\mathcal{PT}$-symmetric mechanical resonators, so that one end of cavity is coupled to the passive mechanical resonator via coupling constant $g_1$, as well as to the active one via $g_2$ as shown in Fig.~\ref{fig1}. The mechanical resonators have equal amounts of loss $(\gamma_1>0)$ and gain $(\gamma_2<0)$, i.e., $\gamma_1=-\gamma_2$, and they are coupled to each other via a mechanical coupling constant $\mu$. The cavity is driven by a strong control laser with angular frequency $\omega_l$ and amplitude $\varepsilon_l$, as well as by a weak probe laser with $\omega_p$ and $\varepsilon_p$. Laser powers can be obtained using $P_i=\hbar \omega_i\varepsilon_i^2$, $i=l,p$, where $\varepsilon_i^2$ is expressed in the frequency dimension.

The Hamiltonian in the rotating frame of the control laser at angular frequency $\omega_l$ is
\begin{eqnarray}
\label{Hh}
\nonumber
\hat{H} & = & \hbar \Delta \hat{a}^\dagger \hat{a} + \hbar \omega_m (\hat{b}_1^\dagger \hat{b}_1 + \hat{b}_2^\dagger \hat{b}_2)-\hbar \mu(\hat{b}_1^\dagger \hat{b}_2 + \hat{b}_2^\dagger \hat{b}_1)\\
& & -\hbar \hat{a}^\dagger \hat{a}g_1( \hat{b}_1^\dagger + \hat{b}_1) -\hbar  \hat{a}^\dagger \hat{a}(g_2 \hat{b}_2^\dagger + g_2^* \hat{b}_2) \\
\nonumber
& &+ i \hbar \sqrt{\eta \kappa} \varepsilon_l (\hat{a}^\dagger-\hat{a}) + i\hbar \sqrt{\eta \kappa}\varepsilon_p (\hat{a}^\dagger e^{-i \omega t}-\hat{a} e^{i \omega t}),
\end{eqnarray}
where $\hat{a}$ ($\hat{a}^\dagger$), and $\hat{b}_1$ ($\hat{b}_1^\dagger$) and $\hat{b}_2$ ($\hat{b}_2^\dagger$) are the annihilation (creation) operators of the cavity and mechanical modes, respectively, $\kappa$ is the cavity decay rate, and $\eta$ is the cavity coupling parameter \cite{weis2010optomechanically}. Here the detuning between the cavity and the control laser is  ${\Delta=\omega_{cav}-\omega_l}$ and that of the probe is $\omega=  \omega_p-\omega_l$ \cite{wang2019mechanical}; the former governs the physics and the latter serves as the primary characterization variable. 
In Eq.~(\ref{Hh}) and in the remainder of this analysis, without loss of generality, we take the two coupling coefficients $\mu$ and $g_1$ as real and non-negative while leaving $g_2 = \left| g_2 \right | e^{i\phi_\ell}$ as complex, with $\phi_\ell$ the total closed-loop coupling phase. Details of the corresponding gauge transformation that justifies this are provided in Appendix~A.

The Heisenberg-Langevin equations which characterize the time evolution of the photon and phonon modes are found based on the above Hamiltonian as 
\begin{eqnarray}
\frac{d\hat{a}}{dt} &= & -i \Delta \hat{a} + i \hat{a} g_1(\hat{b}_1 ^\dagger + \hat{b}_1) + i \hat{a} (g_2\hat{b}_2 ^\dagger + g_2^* \hat{b}_2) \\
\nonumber
& & + \sqrt{\eta \kappa} \varepsilon_l+\sqrt{\eta \kappa} \varepsilon_p e^{-i \omega t}- \frac{\kappa}{2} \hat{a},\\
\frac{d\hat{b}_1}{dt} &=&-i\omega_m \hat{b}_1 + i\mu \hat{b}_2 + ig_1 \hat{a}^\dagger \hat{a}- \frac{\gamma_1}{2} \hat{b}_1,\\
\frac{d\hat{b}_2}{dt}& = &-i\omega_m \hat{b}_2 + i\mu \hat{b}_1 + ig_2 \hat{a}^\dagger \hat{a}- \frac{\gamma_2}{2} \hat{b}_2 ,
\end{eqnarray}
 where the dissipation is introduced within the standard Markovian limit \cite{clerk2010}. Being interested in the probe transmission and phase dispersion characteristics, we discard the thermal and quantum fluctuations of the variables by replacing the operators with their mean values $\hat{\aleph}(t)\rightarrow \langle\hat{\aleph}(t)\rangle\equiv \aleph(t)$ \cite{xiong2012}. Thereupon, our analysis essentially focuses on classical phenomena.

The steady-state solution of this set of equations is given by 
\begin{eqnarray}
\Bar{a} &=& \frac{\sqrt{\eta \kappa} \varepsilon_l}{i\Delta + \frac{\kappa}{2}-i g_1(\Bar{b}_1^* + \Bar{b}_1)-i (g_2\Bar{b}_2^* +g_2^*\Bar{b}_2)} ,\\ 
\Bar{b}_1&=& \frac{[i g_1(i\omega_m+\gamma_2/2)-\mu g_2] |\Bar{a}|^2}{(i\omega_m+\gamma_1/2)(i\omega_m+\gamma_2/2)+\mu^2},\\
\Bar{b}_2&=& \frac{ig_2 |\Bar{a}|^2+i \mu \Bar{b}_1}{i\omega_m+\gamma_2/2}.
\end{eqnarray}
The Heisenberg-Langevin equations are linearized around the steady-state values as $\aleph(t) = \Bar{\aleph}+\delta\aleph(t)$ by ignoring the nonlinear terms \cite{wang2019mechanical,xiong2012}, and the following equation of motions for perturbation terms $\delta\aleph(t)$ are found 
\begin{eqnarray}
\frac{d \delta a}{dt} &= &-i \Delta  \delta a - \frac{\kappa}{2}\delta a + i\delta a g_1(\Bar{b}_1^*+ \Bar{b}_1)  \\
\nonumber
& &+i\Bar{a} g_1(\delta b_1^*+ \delta b_1) +i\delta a(g_2\Bar{b}_2^*+g_2^*\Bar{b}_2) \\ 
\nonumber
& & +i\Bar{a}(g_2\delta b_2^*+g_2^*\delta b_2)+\sqrt{\eta \kappa}\varepsilon_p e^{-i \omega t},\\
\label{pertofdeltaa}
\frac{d \delta b_1}{dt}&= &-i\omega_m \delta b_1 -\frac{\gamma_1}{2} \delta b_1 +ig_1(\Bar{a} \delta a^*+\Bar{a}^* \delta a)\\
\nonumber
& & + i \mu \delta b_2,
\label{deltab1}\\
\frac{d \delta b_2}{dt}&=&-i\omega_m \delta b_2 - \frac{\gamma_2}{2} \delta b_2 +ig_2(\Bar{a} \delta a^*+\Bar{a}^* \delta a) \\
\nonumber
& & + i \mu \delta b_1.
\label{deltab2}
\end{eqnarray}

These equations can be solved by imposing the first-order sidebands only (in the rotating $\omega_l$ frame) as 
\begin{eqnarray}
\delta a & = & A_{1+}e^{i\omega t} + A_{1-}e^{-i\omega t} ,\\
\delta b_1 & = & B_{1+}e^{i\omega t} + B_{1-}e^{-i\omega t}, \\
\delta b_2 & = & C_{1+}e^{i\omega t} + C_{1-}e^{-i\omega t} .
\end{eqnarray}
When the frequency $\omega$ becomes resonant with $\omega_m$ the system starts to oscillate coherently and it creates first-order sidebands, i.e., Stokes and anti-Stokes fields with frequencies (in the nonrotating frame) $\omega_l-\omega$ and $\omega_l+\omega$, respectively \cite{Fundamentals-of-OMIT}. 
Under red-detuned pumping (${\Delta=\omega_m}$), the Stokes field is off-resonance with the cavity mode and it is the anti-Stokes field with frequency $\omega_l+\omega$ that falls into the relevant cavity resonance \cite{Fundamentals-of-OMIT,weis2010optomechanically}. The amplitude of the latter is given by
\begin{equation}
A_{1-}=\frac{\sqrt{\eta \kappa }\varepsilon_p}{\Xi(\omega)-|\Bar{a}|^2 \Lambda(1-\Gamma)},
\end{equation}
where
\begin{eqnarray}
\Xi(\omega)&=&i\Delta+\kappa/2-i \omega-i g_1(\Bar{b}_{1}^{*}+\Bar{b}_{1})-i(g_2\Bar{b}_{2}^{*}+g_2^*\Bar{b}_{2}), \nonumber  \\
\Lambda &=& ig_1 \frac{-i g_1 \alpha_2(\omega_m)-\mu g_2^*}{f_2(\alpha_1,\alpha_2)} +ig_1\frac{i g_1 \alpha_2(-\omega_m)-\mu g_2}{f_1(\alpha_1,\alpha_2)} \nonumber \\
\nonumber
& &+ ig_2\frac{-i g_2^* \alpha_1(\omega_m)-\mu g_1}{f_2(\alpha_1,\alpha_2)}+ig_2^*\frac{i g_2 \alpha_1(-\omega_m)-\mu g_1}{f_1(\alpha_1,\alpha_2)} , \\
\Gamma &=& \frac{|\Bar{a}|^2 \Lambda}{\Xi^*(-\omega)+\Lambda |\Bar{a}|^2} \nonumber,
\end{eqnarray}
with
\begin{eqnarray}
\alpha_1(\omega_m)&=& -i\omega-i\omega_m+\frac{\gamma_1}{2},\\
\alpha_2(\omega_m) &=& -i\omega-i\omega_m+\frac{\gamma_2}{2},\\
f_1(\alpha_1,\alpha_2)&=& \alpha_1(-\omega_m) \alpha_2(-\omega_m)+\mu^2,\\
f_2(\alpha_1,\alpha_2)&=& \alpha_1(\omega_m) \alpha_2(\omega_m)+\mu^2 .
\end{eqnarray}

To obtain transmission of the probe, we use the standard input-output relationship ${S_{out}=S_{in}-\sqrt{\eta \kappa}\langle\hat{a}\rangle}$, where $\langle \hat{a}\rangle= \Bar{a}+\delta a$ \cite{weis2010optomechanically,wang2019mechanical}. The input field $S_{in}$ comes from the driving field in the rotating frame as $S_{in}=\varepsilon_l+\varepsilon_p e^{-i\omega t}$. The amplitude of the anti-Stokes field is found from the output field as ${\varepsilon_p-\sqrt{\eta \kappa}A_{1-}}$ where $S_{out}=\varepsilon_l-\sqrt{\eta\kappa}\Bar{a}+(\varepsilon_p-\sqrt{\eta\kappa}A_{1-})e^{-i\omega t}-\sqrt{\eta\kappa}A_{1+}e^{i\omega t}$. Its division by $\varepsilon_p$ gives the probe transmission amplitude as
\begin{equation}
t_p = 1-\frac{\eta \kappa}{\Xi(\omega)-|\Bar{a}|^2 \Lambda(1- \Gamma)}.
\end{equation}
The derivative of the transmission phase dispersion with respect to the probe frequency determines the group delay
\begin{equation}
\tau_g=\frac{d\psi(\omega_p)}{d\omega_p},
\end{equation}
where the phase dispersion is $\psi(\omega_p)=\arg[t_p(\omega_p)]$ \cite{wang2019mechanical,jiang2018tunable}. 
\par
Next we find the stability matrix to analyze where the coupling of the mechanical gain mode to the cavity introduces instability. Linearized Heisenberg-Langevin equations can be cast in a matrix form $\mathbf{\delta\dot{x}}_n= \mathbf{M}_n \mathbf{\delta x}_n(t)+\mathbf{d}_n$, where the subscript $n$ indicates the number of associated dynamical variables (here $n=6$; for a different choice see Appendix~B) given by the vector $\mathbf{\delta x}_6(t)=(\delta a,\delta a^*,\delta b_1, \delta b_1^*,\delta b_2, \delta b_2^*)^T$ and the vector $\mathbf{d}_6=(\sqrt{\eta \kappa}\varepsilon_p e^{-i\omega t},\sqrt{\eta \kappa}\varepsilon_p e^{i\omega t},0,0,0,0)^T$ denotes the driving terms. The explicit form of the stability matrix is given by
\begin{widetext}
\begin{equation} \label{stability matrix}
\mathbf{M}_6=
\begin{pmatrix}
-i\Delta_a-\kappa/2  & 0 & i\Bar{a}g_1 & i\Bar{a}g_1 & i\Bar{a}g_2^* & i\Bar{a}g_2  \\
0 & i\Delta_a-\kappa/2 & -i\Bar{a}^* g_1 & -i\Bar{a}^* g_1 & -i\Bar{a}^* g_2^* & -i\Bar{a}^* g_2  \\
i\Bar{a}^* g_1 & i\Bar{a}g_1 & -i\omega_m-\frac{\gamma_1}{2}&0&i\mu&0\\
-i\Bar{a}^* g_1&-i\Bar{a}g_1 &0&i\omega_m-\frac{\gamma_1}{2}&0&-i\mu\\
i\Bar{a}^* g_2&i\Bar{a}g_2&i\mu&0&-i\omega_m-\frac{\gamma_2}{2}&0\\
-i\Bar{a}^* g_2^*&-i\Bar{a}g_2^*&0&-i\mu&0&i\omega_m-\frac{\gamma_2}{2}
\end{pmatrix},
\end{equation}
\end{widetext}
where 
\begin{eqnarray}
\Delta_a = \Delta-g_1 (\Bar{b}_1^*+\Bar{b}_1)-g_2 \Bar{b}_2^*-g_2^*\Bar{b}_2\, .
\end{eqnarray}
The system becomes stable when all eigenvalues of the matrix $\mathbf{M}_6$ have negative real parts \cite{xie2018optically,kong2019magnetically}. 

Finally, as we derive under certain approximations in Appendix~B, for the given $g_1$, $g_2$ and other parameters the  intermechanical coupling constant $\mu$, which places the optomechanical system right on the EP, is governed by the analytical expression
\begin{equation}
\mu_\mathrm{EP} \simeq \sqrt{\left|\frac{2\Bar{a}^2 g_1g_2}{\kappa}\right |^2+\bigg[\frac{|\Bar{a}|^2(g_1^2-|g_2|^2)}{\kappa}+\frac{\gamma_1-\gamma_2}{4}\bigg]^2}.
\label{EqmuEP}
\end{equation}
This analytical expression agrees very well with the exact (numerical) solution for the parameter range of interest (see Fig.~10 in Appendix~B).

\section{Parameter Set}
To investigate both $\mathcal{PT}$-broken and unbroken phases, $\mu$ needs to vary from below to above the EP value $\mu_\mathrm{EP}$. As seen in Eq.~(\ref{EqmuEP}),
$\mu_\mathrm{EP}$ explicitly depends on $g_2$ among other variables. The magnitude of $g_2$ is to be determined using the stability analysis, and its phase plays the main role in the control of OMIT, as will be shown below. To ensure the practical relevance of our work, the common parameter set closely follows two ground-state cooling experiments \cite{chan2011laser,massel2011microwave} and consists of $g_1/2\pi=1$~MHz, $\omega_m/2\pi=3.68$~GHz, $\gamma_1=-\gamma_2=0.5\times10^{-2}\omega_m$, $\kappa=0.1\omega_m$, and the wavelength of the control laser $\lambda=1537$~nm. 
Notably, with this choice of $g_1\ll \kappa$, the optomechanical system operates within the weak-coupling limit. We also remark that we did not elaborate on breaking the balanced gain and loss within the mechanical sector, even though a more optimal choice is highly likely \cite{schonleber2016}. Considering the cavity loss as well, the system has overall loss. However, under a simple gauge transformation, the underlying $\mathcal{PT}$ symmetry can be manifested \cite{ozdemir2019parity}.
The remaining parameters are taken as $\eta=\frac{1}{2}$ and $P_c=7.96$~$\mu$W. The latter directly determines the mean number of photons and phonons in the resonators; it will be reduced fivefold in the slow light discussion. We should note that none of the parameters are critical and a different set serving similar purposes is also conceivable.

\section{Results}
\subsection{Stability}
\begin{figure}[H]
  \centering
  \includegraphics[width=0.5\textwidth]{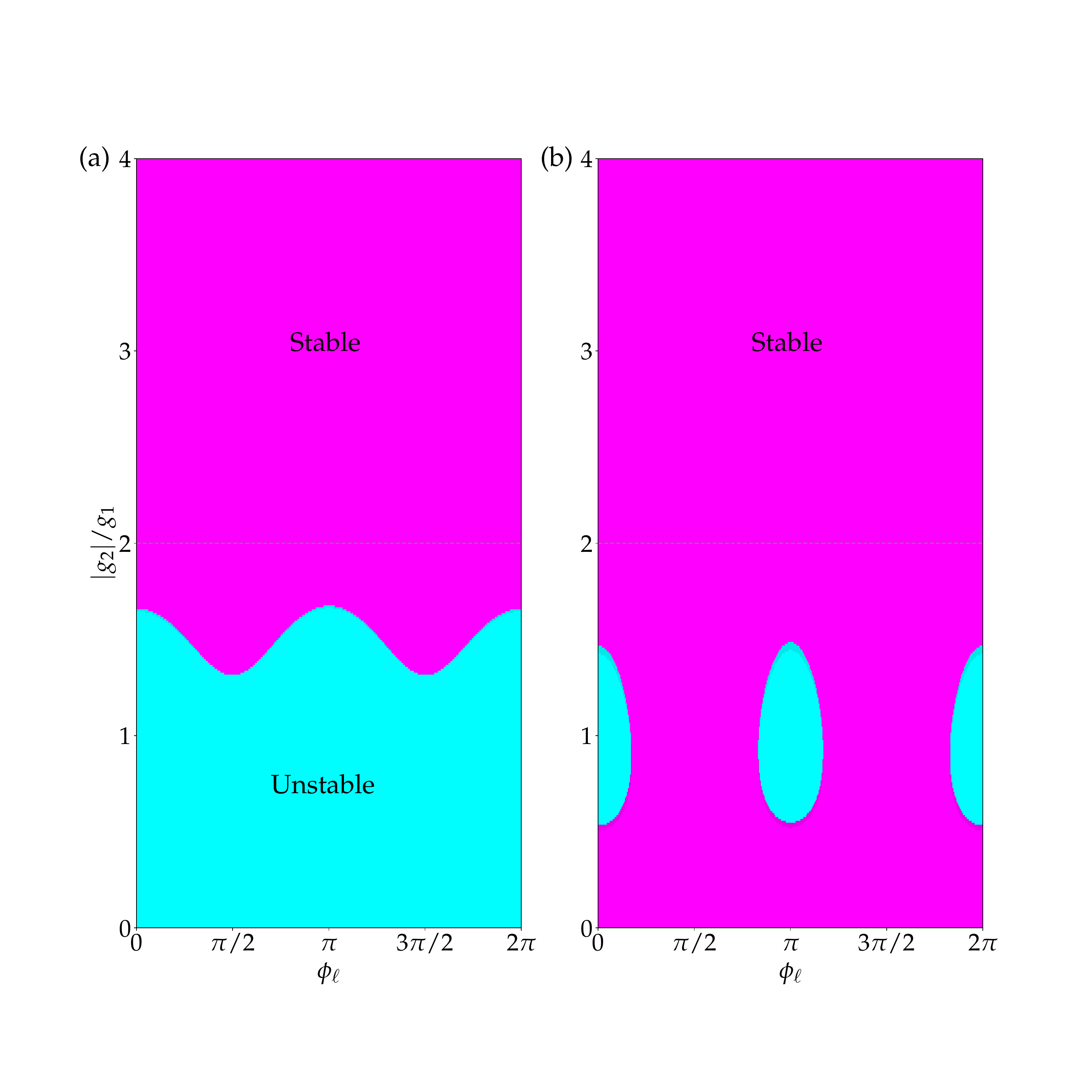}
\caption{Stability analysis as a function of the magnitude and phase of $g_2$ for $g_1/2\pi=1$~MHz and (a) $\mu=0.2(\gamma_1-\gamma_2)$ and (b) $\mu=0.5(\gamma_1-\gamma_2)$. Magenta designates stable, and turquoise unstable regions of the closed-contour interaction. The semitransparent narrow interface region is where the adiabatic elimination estimation (\ref{Eqstab}) disagrees with the exact 6$\times$6 solution. Dashed lines in both panels mark the trajectory used in the following figures. }
\label{fig2}
\end{figure}

First, under the chosen common parameter set we search over the $g_2-\mu$ space to find where the closed-contour interaction  gives a stable response to the anti-Stokes transmission. Solving the eigenvalues of Eq.~(\ref{stability matrix}) numerically, we identify the stable and unstable regions in both $\mathcal{PT}$-broken and -unbroken phases as a function of magnitude and phase of $g_2$ for a fixed coupling constant $g_1$ as displayed in Fig.~\ref{fig2}. There is a slight vertical shift between the estimations based on exact and analytically found eigenvalues [see Appendix~B, in particular Eq.~(\ref{Eqstab})] under adiabatic elimination, as marked by the narrow darker shaded regions. The origin of instability in the system is the active mechanical resonator (gain mode) with $\gamma_2<0$. Increasing its coupling to the lossy cavity via $g_2$ and/or mechanical resonator via $\mu$ (see Fig.~\ref{fig1}) instate the stability. As a matter of fact, in Fig.~\ref{fig2}(b), the limit $|g_2|\rightarrow 0$ becomes unstable, though it is not visible in this scale. Likewise, instability is less prevalent in the $\mathcal{PT}$-unbroken phase ($\mu>\mu_\mathrm{EP}$), as it has a larger intermechanical resonator coupling than the $\mathcal{PT}$-broken phase ($\mu<\mu_\mathrm{EP}$). Based on Fig.~\ref{fig2}, for the remainder of our analysis we choose $|g_2|=2g_1$ (shown by dashed lines) so that the system becomes stable for all values of $\phi_\ell$ in both phases.

\subsection{Root Loci}
\begin{figure}[H]
  \centering
  \includegraphics[width=0.5\textwidth]{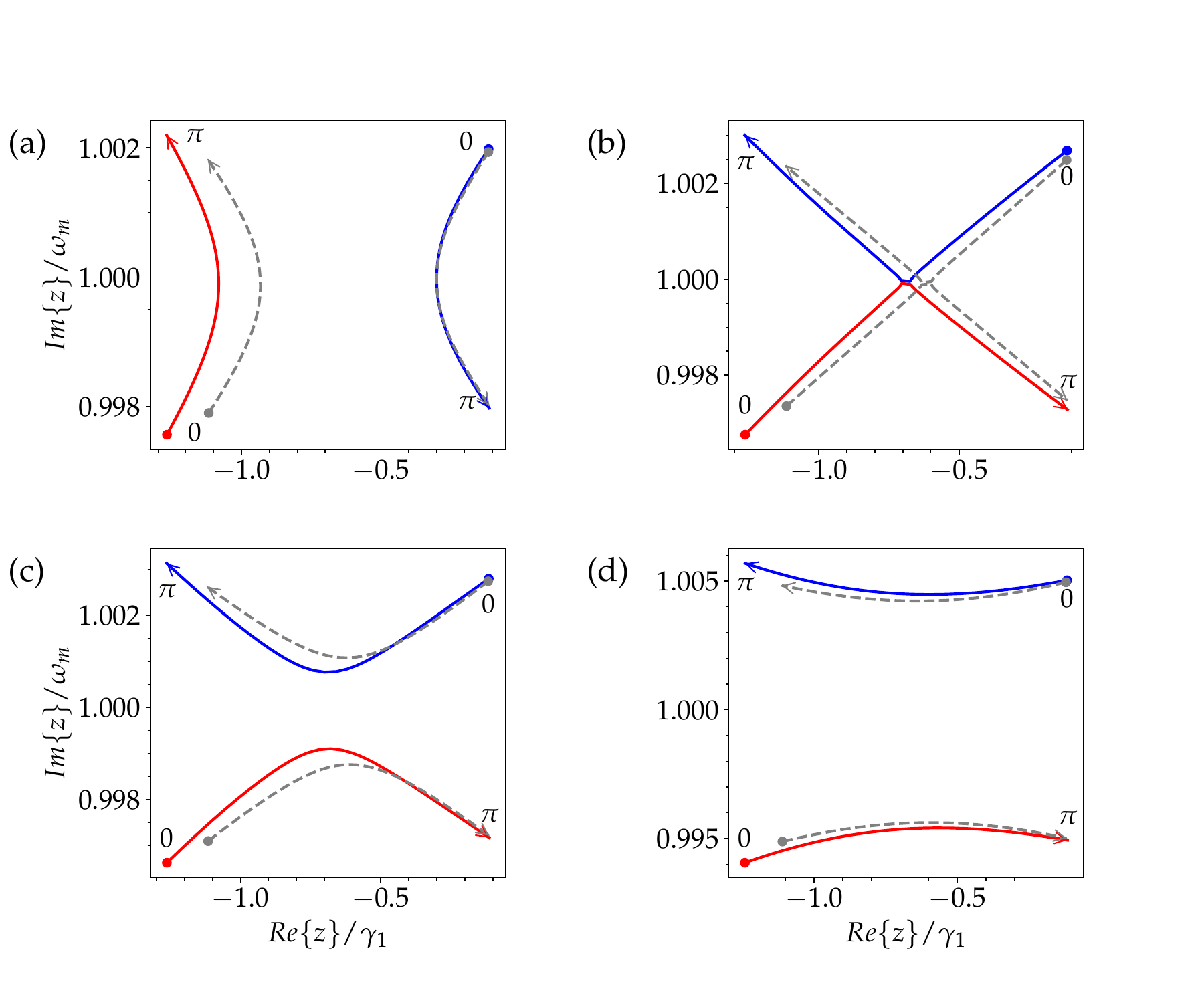}
\caption{Root loci of the mechanical sector eigenvalues as a function of $\phi_\ell\in [0,\pi]$ where the limit values are marked in each panel. Gray dashed lines are the analytically found upper-half-plane eigenvalues as a result of adiabatic elimination [Eq.~(\ref{upperhalfplaneeigenv.})]. The other coupling parameters are $|g_2|=2g_1$ and (a) $\mu/(\gamma_1-\gamma_2)=0.2$, (b) $\mu\simeq\mu_\mathrm{EP}$, (c) $\mu/(\gamma_1-\gamma_2)=0.28$, and (d) $\mu/(\gamma_1-\gamma_2)=0.5$.}
\label{fig3}
\end{figure}

 To further shed light on how to control the behavior of the probe transmission, we study the trajectory of the eigenvalues, also known as the root loci of the system, as a function of $\phi_\ell$ for four different values of $\mu$ chosen below, around, and above the EP. In Fig.~\ref{fig3} we solely track the two roots on the upper half plane originating from the mechanical resonators which display the characteristic traits around the EP. Only the range $\phi_\ell\in [0,\pi]$ is considered, since the roots simply backtrack over $[\pi,2\pi]$. 
 For $\mu<\mu_\mathrm{EP}$, the active and passive mechanical modes largely preserve their individual loss behaviors when $\phi_\ell$ is swept from 0 to $\pi$, while in frequency they traverse in opposite directions from above to below $\omega_m$. For future reference, it needs to be mentioned that $\phi_\ell$ in any case affects both the frequency and the loss of these modes. 
 
 Around the EP, $\mu\sim\mu_\mathrm{EP}$ we see in Fig.~\ref{fig3}(b) a drastic change from the previous vertical migration of the roots as $\phi_\ell$ varies. Coalescence of the eigenvectors occurs, which is the hallmark of $\mathcal{PT}$ symmetry at this special type of the degeneracy point of the eigenvalues  \cite{bender1998real,bender2007making}. Analytical expressions for the eigenvalues [cf. Eq.~(\ref{upperhalfplaneeigenv.})], marked by dashed lines, display trajectories very close to the exact(numerical) ones. However, due to the approximation involved, the analytic $\mu_\mathrm{EP}$ [see Eq.~(\ref{EqmuEP})] slightly differs from the exact value when $|g_2|=2g_1$ and $\phi_\ell=\pi/2$. This is the reason for the deviation in the two sets of eigenvalues in,  e.g., Fig.~\ref{fig3}(b).
 
 Above the EP $\mu>\mu_\mathrm{EP}$ in Figs.~\ref{fig3}(c) and \ref{fig3}(d) the less- and more-lossy modes swap their damping characters as $\phi_\ell$ goes from 0 to $\pi$, so the upper band enhances loss while the lower band acquires relative gain by moving to the right. During this course of character exchange with respect to $\phi_\ell$, midway we expect equal amounts of probe transmission for the $\phi_\ell=\pi/2$ and $3\pi/2$ cases, as will be verified in the transmission spectra. 
 
\subsection{Single and Double OMIT}
\begin{figure}[H]
  \centering
  \includegraphics[width=0.5\textwidth]{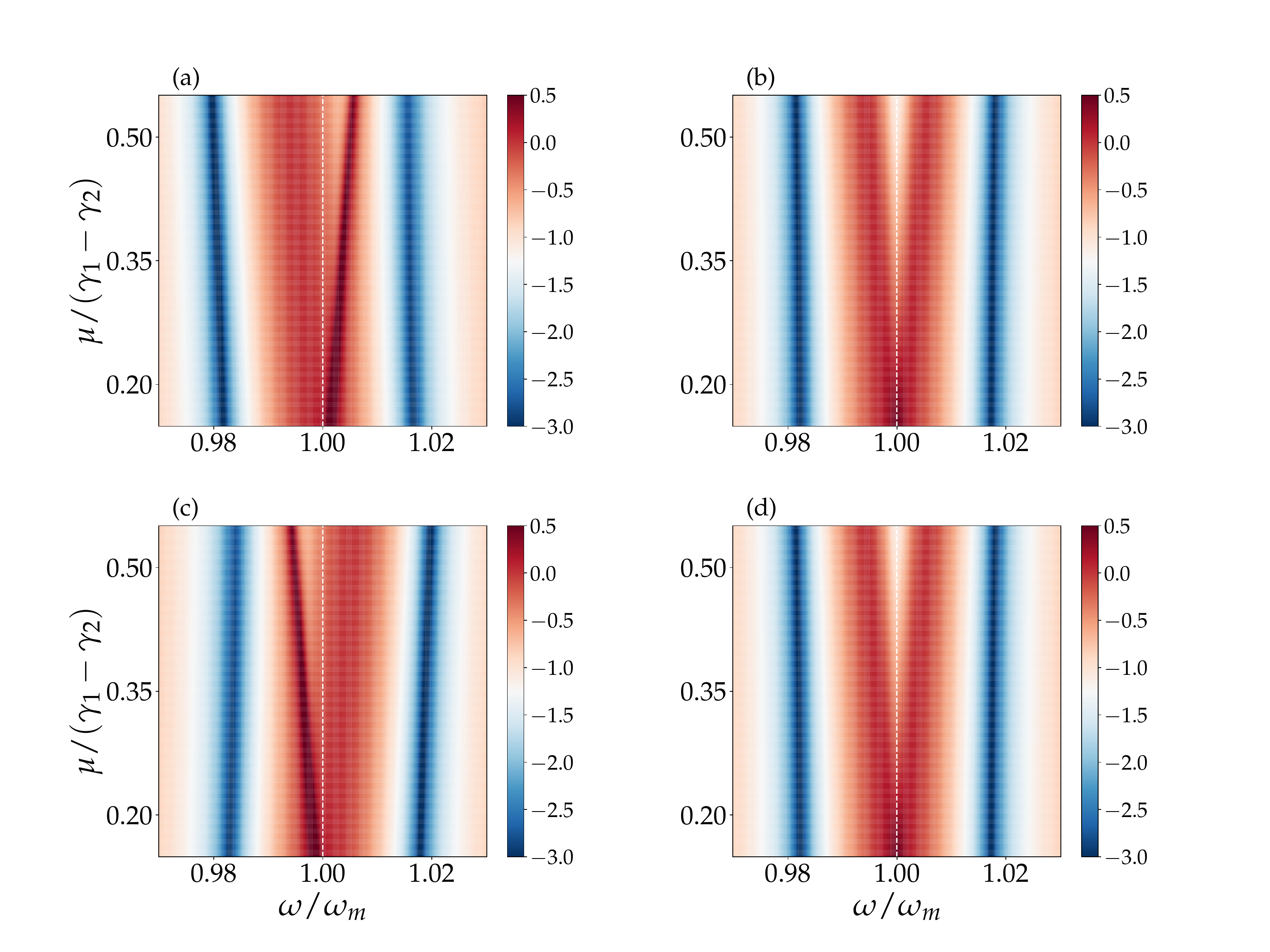}
\caption{Plot of $\mathrm{log}_{10}$ of transmission amplitudes as a function of probe detuning (normalized to $\omega_m$) versus mechanical coupling coefficient $\mu$ (normalized to $\gamma_1-\gamma_2$) for $|g_2|/2\pi=2g_1/2\pi=2$~MHz and (a) $\phi_\ell=0$, (b) $\phi_\ell=\pi/2$, (c) $\phi_\ell=\pi$, and (d) $\phi_\ell=3\pi/2$.}
\label{fig4}
\end{figure}

In the light of the analysis of the root loci, we examine the two-dimensional probe transmission spectrum as a function of $\omega$ and $\mu$. In Fig.~\ref{fig4} this is plotted for four different values of $\phi_\ell$.
First, setting $g_2$ real ($\phi_\ell=0$) yields an asymmetrical spectrum which favors the main transmission in the upper band ($\omega>\omega_m$). The symmetry in the spectrum can be restored by purely imaginary $g_2$ (i.e., $\phi_\ell=\pi/2$ and $3\pi/2$), which also clearly displays the transition from single to double OMIT when $\mu$ changes from $\mu<\mu_\mathrm{EP}$ to $\mu>\mu_\mathrm{EP}$, respectively. The lower band transmission is enabled for $\phi_\ell=\pi$. In this way the transmission can be steered with $\phi_\ell$ by controlling the phase relations and the interference between the direct probe transition and the indirect anti-Stokes field \cite{Fundamentals-of-OMIT}. To ensure that this is a phase-driven effect, we checked that the change in the mean populations of all three resonators remains within 15\% over the full range of $\phi_\ell$ (not shown). Here our interest is focused on controlling the main transmission within the lower and upper bands through the closed-loop phase $\phi_\ell$, whereas previous OMIT studies having photonic-sector $\mathcal{PT}$ symmetry aimed to control the transmission amplitude at a fixed band through the temperature, power of the control field, gain-to-loss ratio, and phase of the phonon pump \cite{lu2019tunable,jiang2018tunable}. 

\begin{figure}[H]
  \centering
  \includegraphics[width=0.5\textwidth]{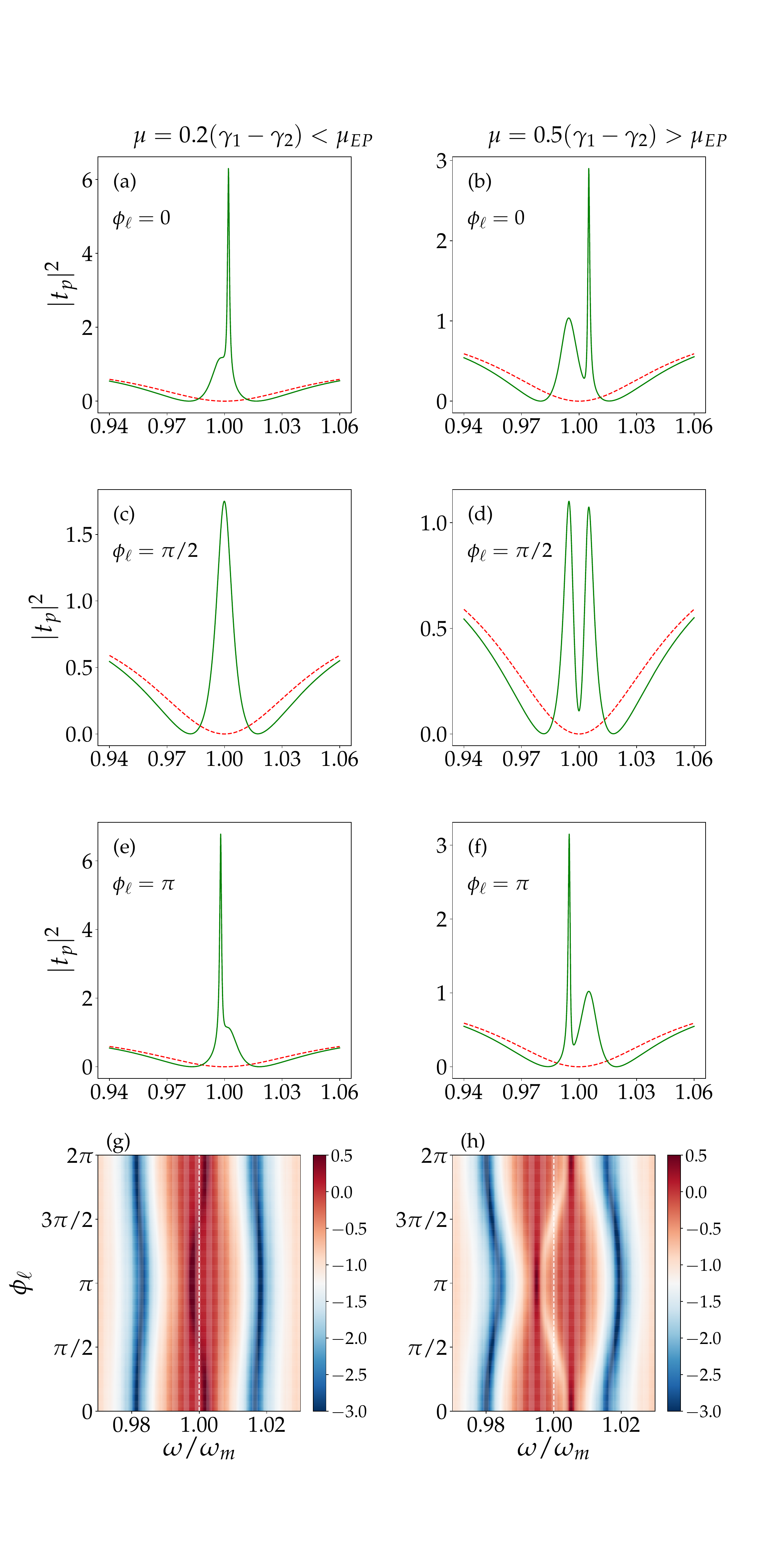}
\caption{Transmission amplitudes as a function of probe detuning (normalized to $\omega_m$) for $|g_2|/2\pi=2g_1/2\pi=2$~MHz in (a), (c), (e), and (g) $\mathcal{PT}$-broken and (b), (d), (f), and (h) $\mathcal{PT}$-unbroken phases.  Red dashed lines correspond to the zero pump case i.e., $\varepsilon_l=0$. Plots (g) and (h) display $\mathrm{log}_{10}$ of transmission amplitudes as a function of the phase of $g_2$ and probe detuning (normalized to $\omega_m$).}
\label{fig5}
\end{figure}

To gain better insight, we examine separately the $\mathcal{PT}$-broken and -unbroken phases. Starting with the former, in Figs.~\ref{fig5}(a), \ref{fig5}(c), and \ref{fig5}(e) we plot again the probe transmission for the three $\phi_\ell$ values considered above. Recall from Fig.~\ref{fig3}(a) that there are two modes, one more lossy than the other, and they are swapped in frequency while retaining their loss rankings under $\phi_\ell$. Thus, we have a single OMIT at $\omega_m$ having a broad width for both $\phi_\ell=\pi/2$ and $3\pi/2$. The peak can be marginally moved above or below $\omega_m$ with $\phi_\ell$; thereby, even for the $\mathcal{PT}$-broken case there is to some extent control over the transmission of the probe. In the $\mathcal{PT}$-unbroken phase shown in Figs.~\ref{fig5}(b), \ref{fig5}(d), and \ref{fig5}(f), two supermodes, namely, dressed states, are responsible for the double transmission peaks \cite{wang2019mechanical}. This double OMIT is clearly seen in Fig.~\ref{fig5}(d) for $\phi_\ell=\pi/2$, whereas in the $\mathcal{PT}$-broken phase we have a single peak in Fig.~\ref{fig5}(c). In accord with the root loci in Fig.~\ref{fig3}, for above the EP, as the roots develop a frequency gap, the transmission peaks are well separated, and the phase steering of the transmission from the upper to the lower band is well resolved in the $\mathcal{PT}$-unbroken phase as in Figs.~\ref{fig5} (b) and \ref{fig5}(f). We should note here that when $g_2=0$ there is still a transition from single to double OMIT via the EP \cite{wang2019mechanical}; however, this comes without the ability to switch within the lower and upper bands. The dashed lines plotted in Figs.~\ref{fig5}(a)-\ref{fig5}(f)  indicate the zero-pump cases, i.e., $\varepsilon_l=0$, where the direct absorption of the probe peak is observed at $\omega_m$, which simply verifies that it is the nonlinear radiation-pressure interaction that endows both single and double OMIT. Finally, Figs.~\ref{fig5}(g) and \ref{fig5}(h) summarize these in the form of two-dimensional density plots.

In the discussion above we used the phase of $g_2$ as the exemplary spectral control parameter. We also checked that identical results are obtained if instead the intermechanical coupling constant  $\mu=|\mu| e^{i\phi_\mu}$ is taken as complex and $\phi_{\mu}$ is swept from $0$ to $\pi$. This ensures that the phase-dependent control is not specific to the coupling constant $g_2$. As a matter of fact, in Appendix~A we show that individual phases of $g_1,g_2$, and $\mu$ forming the closed-loop interaction can be lumped into any one of them, say, $g_2$, via a gauge transformation. 
In this way it constitutes a synthetic gauge field, where the so-called gauge-invariant phase sum \cite{koch2010time} $\phi_\ell=-\phi_1+\phi_2+\phi_{\mu}$ (see Fig.~\ref{fig1}) gives rise to an observable effect, namely, the spectral tunability in the system. This phase discriminates the counterclockwise and clockwise traversals over the closed contours encountering gain and loss sections in different order, which amounts to the breaking of time-reversal symmetry in the $\mathcal{PT}$-symmetric system \cite{deak2012reciprocity}. Moreover, as noted in Appendix~A, $\phi_\ell=0$ and $\pi$ correspond to cases where the time-reversal symmetry is restored, and the double OMIT spectrum is dominated by one of the bands. 

\subsection{Gain-Bandwidth Product}
\begin{figure}[H]
  \centering
  \includegraphics[width=0.5\textwidth]{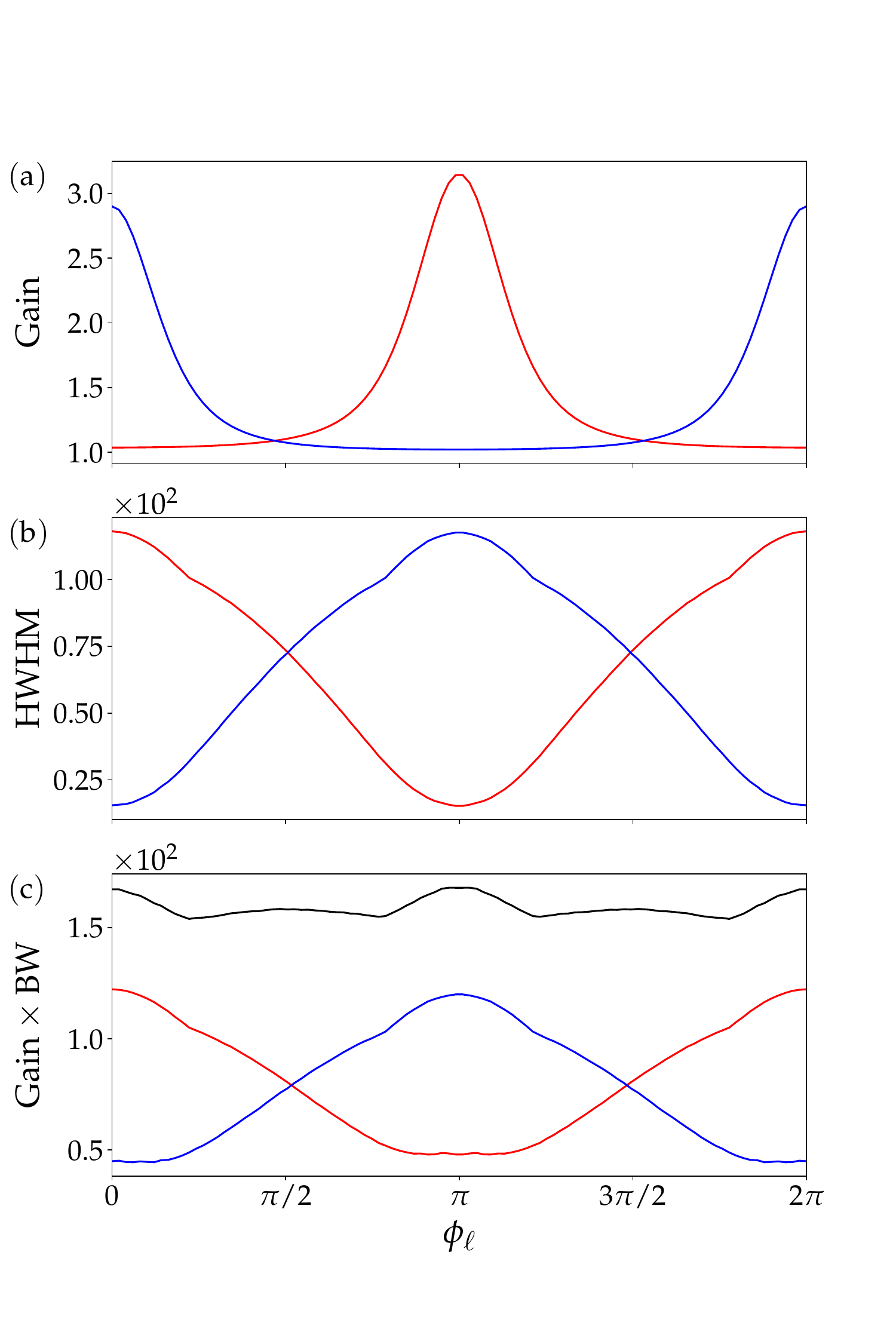}
\caption{Gain-bandwidth product at $\mu=0.5(\gamma_1-\gamma_2)$ in the $\mathcal{PT}$-unbroken phase. The band-width (BW) is in units of $\omega_m$. The lower band ($\omega<\omega_m$) is shown in red and the upper band ($\omega>\omega_m$) in blue. In (c) the black line shows the sum of both bands together.}
\label{fig6}
\end{figure}

\begin{figure}[H]
  \centering
  \includegraphics[width=0.5\textwidth]{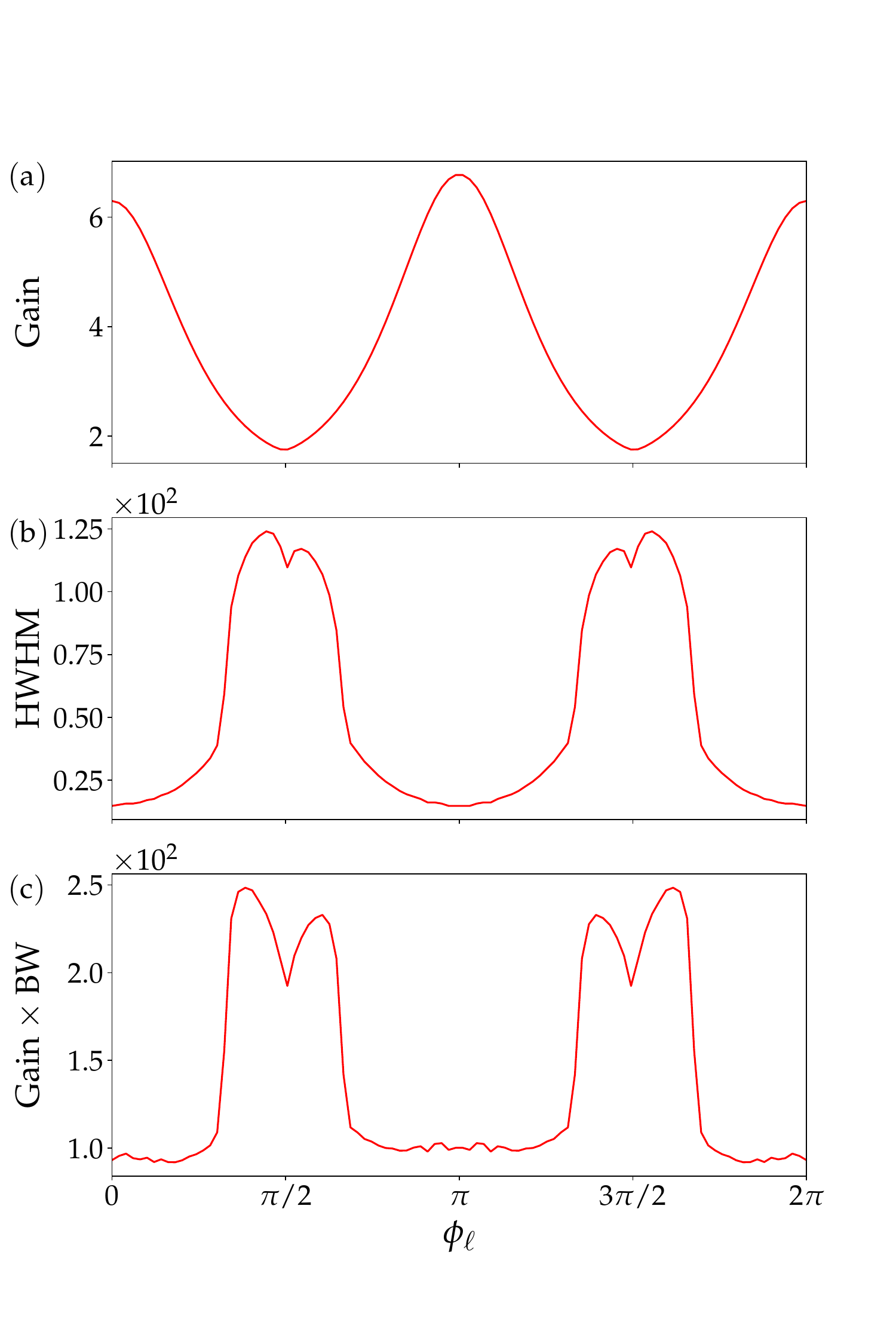}
\caption{Gain-bandwidth product for $\mu=0.2(\gamma_1-\gamma_2)$ in the $\mathcal{PT}$-broken phase. The BW is in units of $\omega_m$.}
\label{fig7}
\end{figure}

Figures~\ref{fig5}(a)-\ref{fig5}(f) also reveal that peak transmission accompanies the narrower of the bands. To investigate this quantitatively, first we consider the gain-bandwidth product in the $\mathcal{PT}$-unbroken phase. Figure~\ref{fig6} displays both the transmission peak and the half-width at half maximum (HWHM) bandwidth for the lower (in red) and upper (in blue) bands separately. The total gain-bandwidth product which accounts for both bands (shown in black) in Fig.~\ref{fig6}(c) remains constant over the full span of the tuning phase $\phi_\ell$. Such a gain-bandwidth trade off is commonly displayed in other optomechanical systems \cite{clerk2010} and optical amplifiers \cite{zhong2020}.
In the case of the $\mathcal{PT}$-broken phase (see Fig.~\ref{fig7}), due to the lack of a clear peak separation, we cannot discriminate between lower and upper bands. Here the inverse behavior of the gain versus the bandwidth is still observed; however, their product rather significantly varies with respect to $\phi_\ell$.

\subsection{Group Delay}
\begin{figure}[H]
  \centering
  \includegraphics[width=0.5\textwidth]{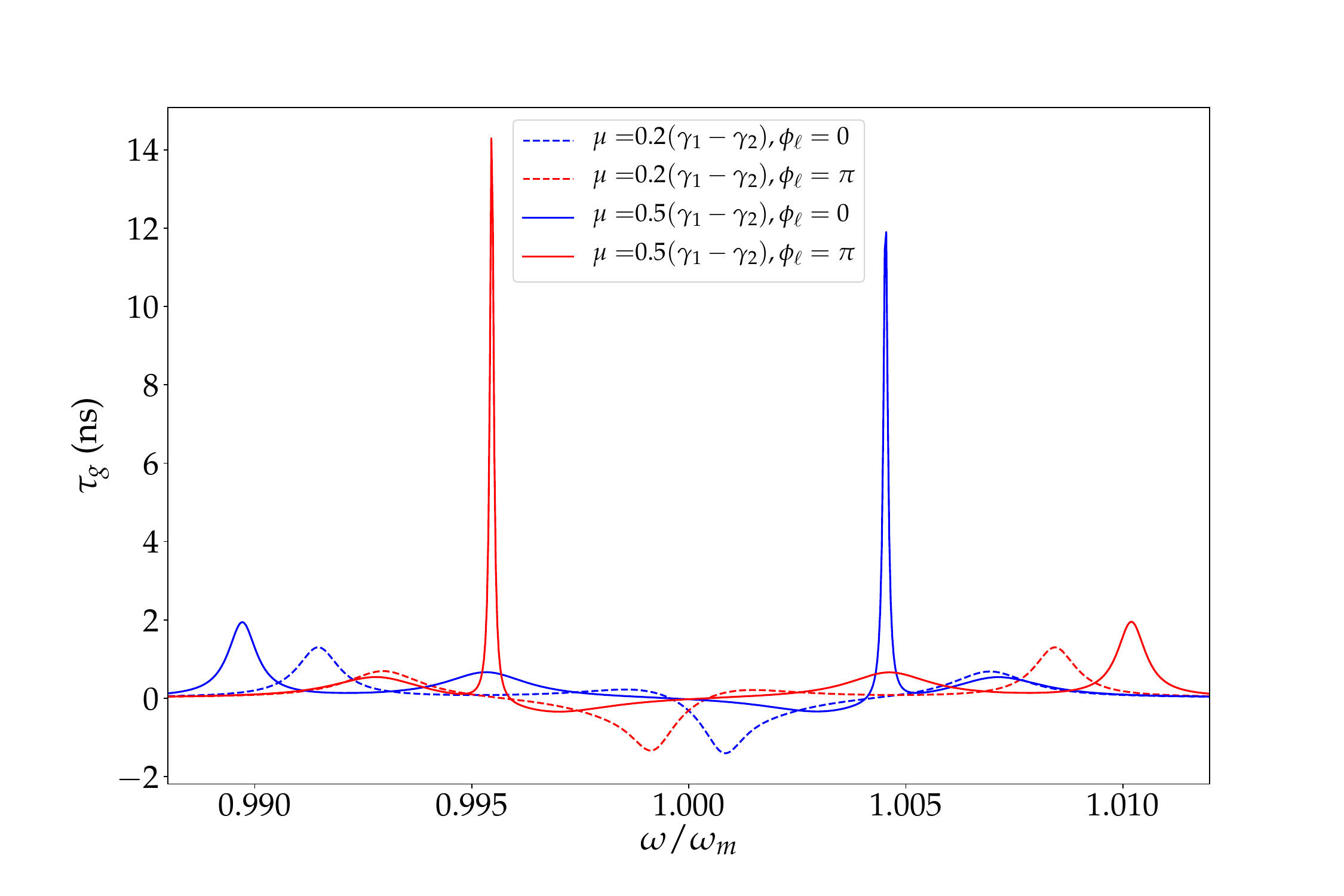}
\caption{Group delay as a function of probe detuning (normalized to $\omega_m$) for $|g_2|/2\pi=2g_1/2\pi=2$~MHz in $\mathcal{PT}$-broken [$\mu=0.2(\gamma_1-\gamma_2)<\mu_\mathrm{EP}$] and, unbroken [$\mu=0.5(\gamma_1-\gamma_2)>\mu_\mathrm{EP}$] phases where $P_c=1.59~\mu$W and $\phi_\ell=0, \pi$. Positive and negative $\tau_g$ correspond to slow and fast light propagation, respectively.}
\label{fig8}
\end{figure}

Next we would like to demonstrate the slow light behavior in the transmission windows of double OMIT in the $\mathcal{PT}$-unbroken phase. To somewhat enhance the effect, in this case we reduce the pump power to $P_c=1.59~\mu$W. 
In Fig.~\ref{fig8}, group delays corresponding to $\phi_\ell=0$ and $\pi$ are plotted as a function of probe detuning in both $\mathcal{PT}$ broken and unbroken phases. The figure clearly reveals that the band in transmission also exhibits a slow light character. Once again, this is continuously tunable by $\phi_\ell$. 
Note that we did not aim to optimize the value of the group delay. Typically, it lies around the submicrosecond range, but as shown theoretically it can be extended to a few milliseconds \cite{liu2017}. We also remark that for the $\mathcal{PT}$-broken phase $\mu<\mu_\mathrm{EP}$, designated with dashed lines in Fig.~\ref{fig8}, a small group \textit{advance} (i.e., fast light) is observed in the pass-bands. 
In comparison, previous studies demonstrated either the same character in both $\mathcal{PT}$-broken and -unbroken phases \cite{zhang2018double} or switching between slow and fast light behaviors by adjusting the gain-to-loss ratio, power of the control field, and the amplitude and phase of the phonon pump \cite{jiang2018tunable}.

\begin{figure}[H]
  \centering
  \includegraphics[width=0.5\textwidth]{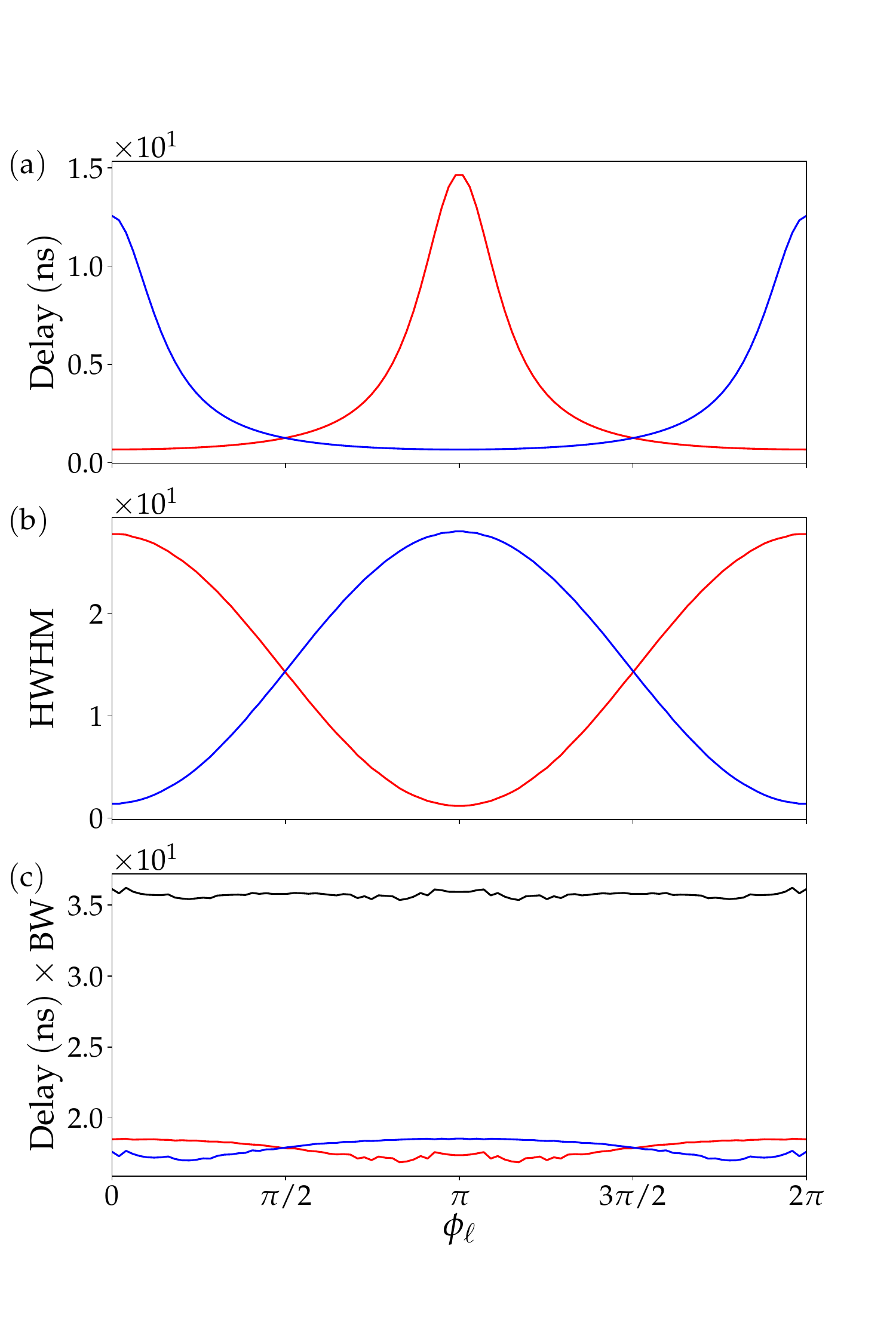}
\caption{Delay-bandwidth product for $|g_2|/2\pi=2g_1/2\pi=2$~MHz, $\mu=0.5(\gamma_1-\gamma_2)>\mu_\mathrm{EP}$ in the $\mathcal{PT}$-unbroken phase, and $P_c=1.59~\mu$W. The BW is in units of $\omega_m$. The lower band ($\omega<\omega_m$) is shown in red, and the upper band ($\omega>\omega_m$) in blue. In (c) the black line shows the sum of both bands together.}
\label{fig9}
\end{figure}

Finally, Fig.~\ref{fig9} illustrates that a delay vs bandwidth trade-off similar to that in the case of gain vs bandwidth applies. This is reminiscent of passive optomechanical systems possessing a constant delay-bandwidth product \cite{bodiya2019}. Because of its insignificant group advance value, the associated bandwidth of the $\mathcal{PT}$-broken phase is not considered here.

\section{Remarks on experimental aspects}
Finally, we would like to discuss some experimental aspects of our theoretical framework.
We begin by recalling that, as mentioned above in Sec.~III, our parameter set is specifically chosen to ensure the experimental feasibility \cite{chan2011laser,massel2011microwave}.
A key concern is how to achieve the required mechanical $\mathcal{PT}$ symmetry in practice. Hitherto, a critical innovation has been the phonon laser, 
in which an originally lossy mechanical mode of, say a microtoroidal, resonator can be brought to the phonon lasing regime by the effective 
mechanical gain induced by optical modes \cite{grudinin2010,jing2014}. 
Unlike a coherent phonon pump, the phonon laser avails modeling the overall cavity with a simple gain term above its transparency \cite{zhang2018phonon}.
This grants a further advantage compared to photonic counterparts, namely in experimentally spotting the EP. Indeed, optical lasing modes have the undesirable susceptibility to 
become unstable in the vicinity of an EP \cite{zyablovsky2016}, making it rather formidable to explore the parameter space around the EP. 
On the other hand the EP associated with the mechanical degrees of freedom offers an easier route to circumvent this problem, 
as experimentally demonstrated using phonon lasers, where for instance, an additional tip-induced loss enables one to steer the phonon laser around the mechanical EP \cite{zhang2018phonon}. We mention yet another proposal for introducing \textit{balanced} gain and loss to two mechanical resonators by driving the associated optomechanical cavities with red- and blue-detuned optical lasers \cite{xu2015mechanical}.

Another crucial aspect of our scheme is the necessity for coupling the two mechanical modes. Due to the rapid progress made in nanoelectromechanical fabrication techniques, this is no longer a technical obstacle \cite{okamoto2013,huang2013,schmid2016,huang2017nonreciprocal}. 
Nevertheless, a simpler alternative to two distinct mechanical resonators is using a \textit{single} square-shaped silicon nitride membrane's 
twofold degenerate vibrational modes, with their coupling being achieved via the radiation pressure when placed in a high finesse optical cavity \cite{xu2016}. The power and the detuning of the driving laser enable one to experimentally map out the complex eigenvalues of the mechanical modes of the membrane by monitoring its heterodyne response signal; The location of the EP is unambiguously resolved in 
agreement with the characteristic features displayed depending on whether the EP is encircled or not \cite{xu2016}. Additionally, our model demands the continuous tunability of the loop coupling phase. This has been experimentally demonstrated, for instance, in a superconducting circuit optomechanical system in which mechanical motion is capacitively coupled to a multimode microwave circuit, where the microwave pump's phase is linked to the coupling phase by a constant offset \cite{bernier2017nonreciprocal}.

As in the case of phonon laser \cite{jing2014,he2016dynamical} and $\mathcal{PT}$-symmetric optomechanical \cite{jing2015optomechanically,jing2017high,liu2017,lu2018,wang2019mechanical} studies, in our model we use a \textit{fixed} gain rate $\gamma_2$. This leaves out gain instability and saturation considerations which have been specifically addressed in photonic \cite{hassan2015,teimourpour2017,he2018transmission,barton2018,sunada2018} and recently in optomechanical $\mathcal{PT}$ systems \cite{xie2020laser,hao2021gain}. 

\section{Conclusions}
The synthetic gauge field concepts have proved to be very fruitful especially in photonics. In this work we demonstrated this on an optomechanical system involving a $\mathcal{PT}$-symmetric mechanical pair of resonators. Introducing a closed-contour interaction in this setting breaks the time-reversal symmetry and imparts tunability through the gauge-invariant phase sum. Specifically working in the stable region, we showed that in the $\mathcal{PT}$-unbroken phase it enables the steering of the transmission and slow light characteristics within the double OMIT bands while keeping the bandwidth products essentially constant. In the $\mathcal{PT}$-broken phase it again provides up to 50\% variation of the OMIT bandwidth. The rationale behind these phenomena can be simply understood by the loci of the mechanical supermodes over the complex plane as a function of the loop coupling phase. This work was confined to the mean properties of the dynamical variables. An interesting extension can be the investigation of synthetic gauge field control in the quantum regime of optomechanical systems \cite{li2017theoretical}. 

\textit{Note added.} Recently, we became aware of the work of Jiang $\textit{et~al}$. based on a similar optomechanical setting reporting the ground-state cooling of mechanical resonators mediated by a phononic gauge field \cite{jiang2021energy}.

\begin{acknowledgments}
We are thankful to C. Y\"{u}ce for fruitful comments.
\end{acknowledgments}

\appendix
\section*{Appendix A: Gauging out individual coupling phases}
\label{appendix:a}
In this appendix we would like to show how the individual coupling phases of $\mu$ and $g_1$
can be gauged out leading to the form in Eq.~(\ref{Hh}) \cite{koch2010time}. We begin by first restoring all the phases of the coupling coefficients
in this Hamiltonian, while dropping the free and the drive terms
\renewcommand{\theequation}{A\arabic{equation}}
\setcounter{equation}{0}

\begin{equation}
\label{Hc}
\hat{H}_c = -\hbar \mu e^{i\phi_\mu} \hat{b}_1^\dagger \hat{b}_2 - \hbar g_1 e^{i\phi_1} \hat{a}^\dagger \hat{a} \hat{b}_1^\dagger 
-\hbar |g_2| e^{i\phi_2} \hat{a}^\dagger \hat{a} \hat{b}_2^\dagger + \mbox{H.c.},
\end{equation}
where $\mu,g_1\in\mathbb{R}$ and $\mbox{H.c.}$ stands for the Hermitian conjugate. The phase of the photonic cavity mode operator 
$\hat{a}$ has no importance in $\hat{H}_c$. We can cancel the phases $\mu$ and $g_1$ by the following gauge transformation of the mechanical mode operators:
\begin{eqnarray}
\hat{b}_1^\dagger & \rightarrow & \hat{b}_1^\dagger e^{-i\phi_1},\\
\hat{b}_2^\dagger & \rightarrow & \hat{b}_2^\dagger e^{i\phi_\mu-i\phi_1}.
\end{eqnarray}
This transforms Eq.~(\ref{Hc}) to
\begin{equation}
\hat{H}_c = -\hbar \mu \hat{b}_1^\dagger \hat{b}_2 - \hbar g_1 \hat{a}^\dagger \hat{a} \hat{b}_1^\dagger 
-\hbar |g_2| e^{i\phi_\ell} \hat{a}^\dagger \hat{a} \hat{b}_2^\dagger + \mbox{H.c.},
\end{equation}
where $\phi_\ell=-\phi_1+\phi_2+\phi_\mu$, so that we obtain the field coupling terms of Eq.~(\ref{Hh}) by defining $g_2\equiv |g_2| e^{i\phi_\ell}$. Thus, we can gauge out the individual coupling phases, and only a single closed-loop overall phase, $\phi_\ell$ remains. Also note that time reversal symmetry is attained for $\phi_\ell\in 0,\pi$.

\section*{Appendix B: Adiabatic elimination of the cavity mode}
\label{appendix:b}
\renewcommand{\theequation}{B\arabic{equation}}
\setcounter{equation}{0}
The analytical solution of the characteristic polynomial of 6$\times$6 stability matrix (Eq.~(\ref{stability matrix})) is not possible. Therefore, we first eliminate the cavity modes adiabatically and reduce the matrix size to 4$\times$4.  We owe this adiabatic elimination approximation to the fact that the cavity decay rate is much greater than the loss and gain of mechanical oscillators (${\kappa \gg |\gamma_{1,2}|}$)  \cite{xu2015mechanical}. We begin by integrating Eq.~(\ref{pertofdeltaa}) by ignoring the $\delta b_{i}$ and $\delta b_{i}^*$ dependences and dropping the probe excitation term since it has no effect on the stability or root loci of the system, yielding 
\begin{widetext}
\begin{eqnarray}
\delta a(t) & = & e^{-(\kappa/2+i\Delta_a)t}\bigg[\delta a(0)+ \int_{0}^{t} i\Bar{a}(g_1\delta b_1^*+g_1\delta b_1 + g_2\delta b_2^*+g_2^*\delta b_2)e^{(\kappa/2+i\Delta_a)\tau} d\tau\bigg].
\label{delta}
\end{eqnarray}
Next, assuming that $\delta b_i(t)$ and $\delta b_i^*(t)$ are not affected by $\delta a$ for $t \gg \kappa^{-1}$, we obtain
\begin{eqnarray}
\delta b_i(\tau) &=& \delta b_i(0)e^{-(\gamma_i/2+i\omega_m)\tau}, 
\label{deltab}\\
\delta b_i^*(\tau) &=& \delta b_i^*(0)e^{-(\gamma_i/2-i\omega_m)\tau}.
\label{deltab22}
\end{eqnarray}
Inserting Eqs.~(\ref{deltab}) and (\ref{deltab22}) into Eq.~(\ref{delta}) and carrying out the integration, we get
\begin{eqnarray}
\delta a(t) &=& i\Bar{a}\bigg[\frac{g_1 \delta b_1^*(t)}{(\frac{\kappa}{2}-\frac{\gamma_1}{2})+i(\Delta_a+\omega_m)}+ \frac{g_1 \delta b_1(t)}{(\frac{\kappa}{2}-\frac{\gamma_1}{2})+i(\Delta_a-\omega_m)}\label{delt.}\\
\nonumber
&&+\frac{g_2 \delta b_2^*(t)}{(\frac{\kappa}{2}-\frac{\gamma_2}{2})+i(\Delta_a+\omega_m)}+\frac{g_2^* \delta b_2(t)}{(\frac{\kappa}{2}-\frac{\gamma_2}{2})+i(\Delta_a-\omega_m)}\bigg].
\end{eqnarray}

The next step is to insert Eq.~(\ref{delt.}) and its complex conjugate into  Eqs.~(\ref{deltab1}) and (\ref{deltab2}) to eliminate the $\delta a$ terms. Dropping the $\gamma_{1,2}$ terms compared to $\kappa$ in the denominators we get
\begin{eqnarray}
\frac{\delta b_1}{dt} &=& -(\frac{\gamma_1}{2}+i\omega_m)\delta b_1+i\mu \delta b_2 + \frac{i2\Delta_a |\Bar{a}|^2}{\frac{\kappa^2}{4}+\Delta_a^2-\omega_m^2-i\kappa \omega_m}(\delta b_1 g_1^2+ \delta b_2 g_1 g_2^*)\\
\nonumber
&&+\frac{i2\Delta_a |\Bar{a}|^2}{\frac{\kappa^2}{4}+\Delta_a^2-\omega_m^2+i\kappa \omega_m}(\delta b_1^* g_1^2+ \delta b_2^* g_1 g_2),\\
\frac{\delta b_2}{dt} &=& -(\frac{\gamma_2}{2}+i\omega_m)\delta b_2+i\mu \delta b_1 + \frac{i2\Delta_a |\Bar{a}|^2}{\frac{\kappa^2}{4}+\Delta_a^2-\omega_m^2-i\kappa \omega_m}(\delta b_2 |g_2|^2+ \delta b_1 g_2 g_1)\\
\nonumber
&&+\frac{i2\Delta_a |\Bar{a}|^2}{\frac{\kappa^2}{4}+\Delta_a^2-\omega_m^2+i\kappa \omega_m}(\delta b_2^* g_2^2+ \delta b_1^* g_2 g_1),\\
\frac{\delta b_1^*}{dt} &=& -(\frac{\gamma_1}{2}-i\omega_m)\delta b_1^*-i\mu \delta b_2^* - \frac{i2\Delta_a |\Bar{a}|^2}{\frac{\kappa^2}{4}+\Delta_a^2-\omega_m^2+i\kappa \omega_m}(\delta b_1^* g_1^2+ \delta b_2^* g_1 g_2)\\
\nonumber
&&-\frac{i2\Delta_a |\Bar{a}|^2}{\frac{\kappa^2}{4}+\Delta_a^2-\omega_m^2-i\kappa \omega_m}(\delta b_1 g_1^2+ \delta b_2 g_1 g_2^*),\\
\nonumber
\frac{\delta b_2^*}{dt} &=& -(\frac{\gamma_2}{2}-i\omega_m)\delta b_2^*-i\mu \delta b_1^* - \frac{i2\Delta_a |\Bar{a}|^2}{\frac{\kappa^2}{4}+\Delta_a^2-\omega_m^2+i\kappa \omega_m}(\delta b_2^* |g_2|^2+ \delta b_1^* g_2^* g_1)\\
&&-\frac{i2\Delta_a |\Bar{a}|^2}{\frac{\kappa^2}{4}+\Delta_a^2-\omega_m^2-i\kappa \omega_m}(\delta b_2 {g_2^*}^2+ \delta b_1 g_2^* g_1).\\
\nonumber
\end{eqnarray}
This set of equations can be cast into a 4$\times$4 stability matrix form $\mathbf{\delta\dot{x}}_4=\mathbf{M}_4 \mathbf{\delta x}_4$ where $\mathbf{\delta x}_4=[\delta b_1, \delta b_2, \delta b_1^*,\delta b_2^*]^T$ as
\begin{equation} \label{stability matrix2}
\mathbf{M}_4=
\begin{pmatrix}
-i\omega_m-\frac{\gamma_1}{2}+iQ^*g_1^2  & i(Q^*g_1g_2^*+\mu) & iQg_1^2 & iQg_1g_2 \\ i(Q^*g_1g_2+\mu) &
-i\omega_m-\frac{\gamma_2}{2}+iQ^*|g_2|^2 & iQg_1g_2 & iQg_2^2\\  
-iQ^*g_1^2 & -iQ^* g_1g_2^*& i\omega_m-\frac{\gamma_1}{2}-iQg_1^2&-i(Qg_1g_2+\mu)\\
-iQ^*g_1g_2^*&-iQ^*{g_2^*}^2 &-i(Qg_1g_2^*+\mu)&i\omega_m-\frac{\gamma_2}{2}-iQ|g_2|^2\\
\end{pmatrix},
\end{equation}
with $Q\equiv \frac{2\Delta_a|\Bar{a}|^2}{\frac{\kappa^2}{4}+\Delta_a^2-\omega_m^2+i\kappa\omega_m}$.
A highly instrumental approximation is to neglect the coupling between conjugate variable pairs, which leads to the 2$\times$2 block-diagonal form
\begin{equation} \label{stability matrix3}
\mathbf{M}_4 \simeq
\begin{pmatrix}
-i\omega_m-\frac{\gamma_1}{2}+iQ^*g_1^2  & i(Q^*g_1g_2^*+\mu) & 0 & 0 \\ 
i(Q^*g_1g_2+\mu) &
-i\omega_m-\frac{\gamma_2}{2}+iQ^*|g_2|^2 & 0 & 0\\  
0 & 0& i\omega_m-\frac{\gamma_1}{2}-iQg_1^2&-i(Qg_1g_2+\mu)\\
0&0 &-i(Qg_1g_2^*+\mu)&i\omega_m-\frac{\gamma}{2}-iQ|g_2|^2\\
\end{pmatrix}.
\end{equation}
The upper half-plane eigenvalues can be solved as 
\begin{eqnarray}
\lambda_{1,2}=\frac{\lambda_{m1}+\lambda_{m2}}{2}\pm \bigg[\frac{(\lambda_{m1}+\lambda_{m2})^2}{4}-(\lambda_{m1}\lambda_{m2}+P)\bigg]^{1/2},
\label{upperhalfplaneeigenv.}
\end{eqnarray}
\end{widetext}
where
\begin{eqnarray}
\lambda_{m1} &=&i(\omega_m-Qg_1^2)-\frac{\gamma_1}{2},\\
\lambda_{m2} &=&i(\omega_m-Q|g_2|^2)-\frac{\gamma_2}{2},\\
P &=& (Qg_1g_2+\mu)(Qg_1g_2^*+\mu).
\end{eqnarray}

At the EP, eigenvalues coalesce, and for this we set $\phi_\ell=\frac{\pi}{2}$ so that $g_2=i|g_2|$. Choosing the positive root we get
\begin{equation}
\mu_\mathrm{EP} = \left\{\left[\frac{iQ}{2}(g_1^2-|g_2|^2)+\frac{\gamma_1-\gamma_2}{4}\right]^2-(Qg_1g_2)^2\right\}^{1/2}\,.
\end{equation}
Note that for $\Delta_a \approx \omega_m$ we can make another approximation $Q \simeq -\frac{i2|\Bar{a}|^2}{\kappa}$, which renders $\mu_\mathrm{EP}\in \mathbb{R}$, and we obtain 
\begin{equation}
\mu_\mathrm{EP} \simeq \sqrt{\left|\frac{2\Bar{a}^2 g_1g_2}{\kappa}\right |^2+\bigg[\frac{|\Bar{a}|^2(g_1^2-|g_2|^2)}{\kappa}+\frac{\gamma_1-\gamma_2}{4}\bigg]^2}.
%\label{EqmuEP}
\end{equation}

\begin{figure}[H]
  \centering
  \includegraphics[width=0.5\textwidth]{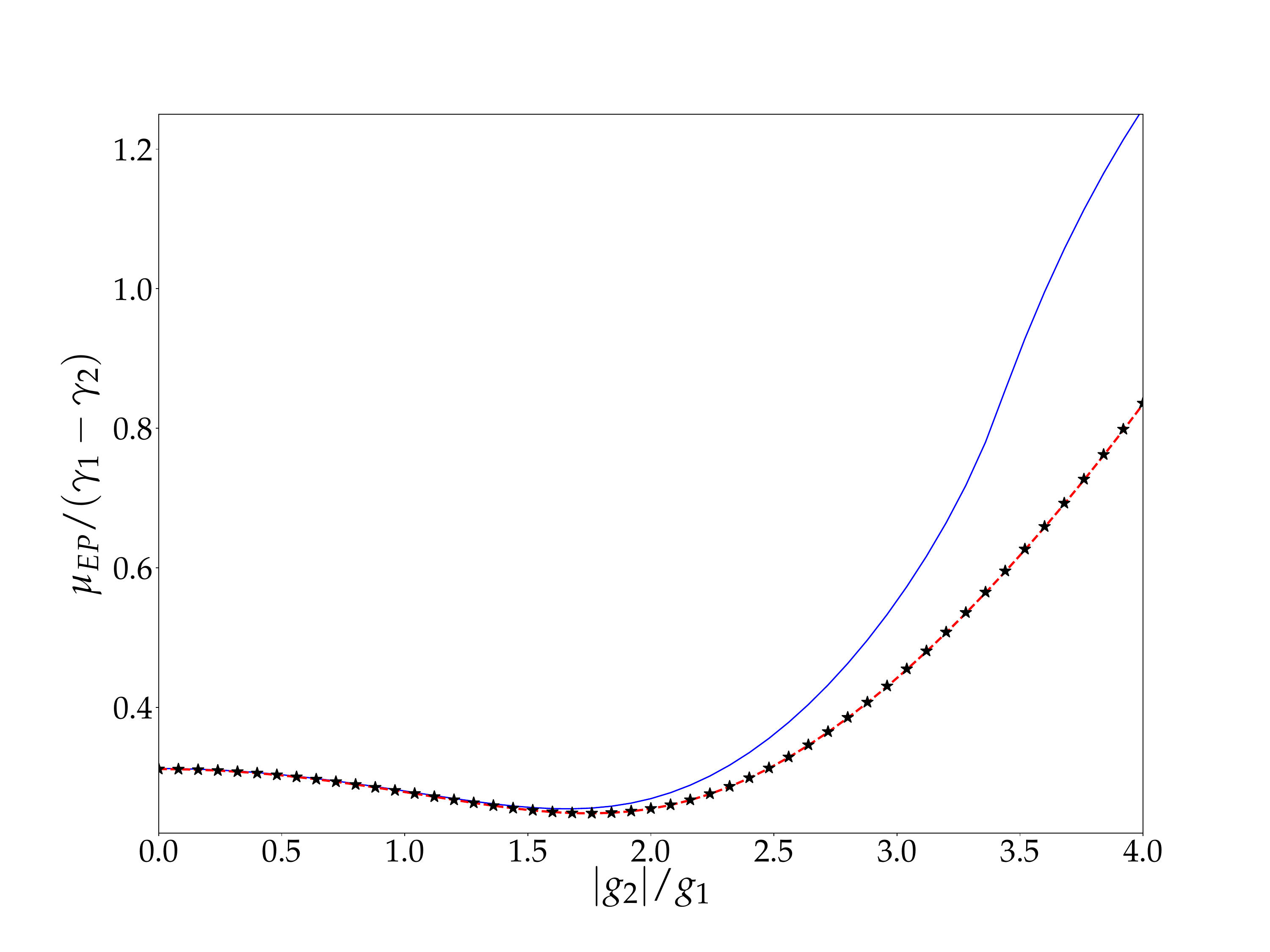}
\caption{Comparison of analytic and numerical $\mu_\mathrm{EP}$ (normalized to $\gamma_1-\gamma_2$) as a function of magnitude of $g_2$. Blue solid, red dashed and black star marked lines correspond to the exact 6$\times$6 case (numerical) case, the 4$\times$4 adiabatic elimination (numerical) and the 4$\times$4 block diagonal (analytic) adiabatic elimination case with further approximations, respectively. }
\label{fig10}
\end{figure}
Figure~\ref{fig10} compares the $\mu_\mathrm{EP}$ under the exact(numerical) and approximate(analytic) treatments. They can be observed to be in very good agreement up to $|g_2|=2g_1$, which is the value used in showcasing our results. \par   
For the stability condition of the optomechanical system an analytical expression is obtained within the validity of the aforementioned approximations,
\begin{equation}
\label{Eqstab}
\operatorname{Re}(\lambda_{1,2})<0,
\end{equation}
where 
\begin{equation}
\label{Eqanaleig}
\lambda_{1,2}\simeq \frac{\lambda_{m1}+\lambda_{m2}}{2} \pm \frac{\lambda_{m1}-\lambda_{m2}}{2} \sqrt{1-\frac{4P}{(\lambda_{m1}-\lambda_{m2})^2}}.
\end{equation}
Here, $Q\equiv \frac{2\Delta_a|\Bar{a}|^2}{\frac{\kappa^2}{4}+\Delta_a^2-\omega_m^2+i\kappa\omega_m}$ needs to be used in obtaining $P$ because its approximate form ($Q \simeq -\frac{i2|\Bar{a}|^2}{\kappa}$) does not perform as well as the one shown in Fig.~2.

\end{document}